\documentclass[%
reprint,superscriptaddress,
 amsmath,amssymb,
 pre,
 aps,
 showkeys,footinbib
]{revtex4-2}
\usepackage{hyperref}
\usepackage{graphicx}
\usepackage{amsfonts} 
\usepackage{thmbox} 
\usepackage{empheq} 
\usepackage{version}
\usepackage{times}
\usepackage{makeidx}
\usepackage{enumerate}
\usepackage{amsmath}
\usepackage{fancyhdr}
\usepackage{color}

\newcommand{\beq}{\begin{equation}}     \newcommand{\eeq}{\end{equation}}
\newcommand{\beqa}{\begin{eqnarray}}    \newcommand{\eeqa}{\end{eqnarray}}
\newcommand{\bde}{\begin{description}}  \newcommand{\ede}{\end{description}}
\newcommand{\ben}{\begin{enumerate}}    \newcommand{\een}{\end{enumerate}}
             
\newcommand{\noi}{\noindent\mbox{}}

\newcommand{\la}{\langle}               \newcommand{\ra}{\rangle}

\newcommand{\intOI}{{\int^{+\infty}_{0}}}

\newcommand{\N}{{\mathbb{N}}} 

\newcommand{\inv}[1]{{\frac{1}{#1}}}

\newcommand{\inRbracket}[1]{{\left({#1}\right)}}
\newcommand{\inSbracket}[1]{{\left[{#1}\right]}}

\newtheorem[L]{thm}{Theorem}[section]
\newtheorem{cor}[thm]{Corollary}
\newtheorem{theorem}{{\sf Assertion :}}[section] 
\newtheorem{definition}{\sf Definition} 
 
\newtheorem{lemma}[theorem]{Lemma}
\newcommand{\bth}{\begin{thm}}  
\newcommand{\blem}{\begin{lemma}}  
\newcommand{\elem}{\end{lemma}}  
\newcommand{\bpr}{\begin{proof}}  
\newcommand{\epr}{\end{proof}} 
\newcommand{\bdefine}{\begin{definition}}  
\newcommand{\edefine}{\end{definition}}  
\newcommand{\bcor}{\begin{cor}} 
\newcommand{\ecor}{\end{cor}}  
\newcommand{\bprop}{\begin{example}[Property]}  
\newcommand{\eprop}{\end{example}}  
\newcounter{formulaire}
\newcommand{\beqf}{\addtocounter{formulaire}{1}\begin{equation}}
\newcommand{\eeqf}{\tag{R \arabic{formulaire}}\end{equation}}
\newcommand{\beqaf}{\addtocounter{formulaire}{1}\begin{equation}\begin{array}{rcl}}
\newcommand{\eeqaf}{\end{array}\tag{R \arabic{formulaire}}\end{equation}}

\newcommand{\hattau}{{\hat{\tau}}}

\usepackage{dcolumn}
\usepackage{subfigure}
\usepackage{listings} 

\begin{document}

\title{Derivation of the First Passage Time Distribution for Markovian Process on Discrete Network }
\author{Ken Sekimoto}
\affiliation{Laboratoire Gulliver, UMR CNRS 7083, ESPCI Paris, Universit\'e PSL
10 rue Vauquelin, 75005, Paris, France}
\affiliation{Mati\`eres et Systm\`emes Complexes, UFR de Physique, Universit\'e Paris Cit\'e, 75013, Paris, France }
\email[Corresponding author: ]{ken.sekimoto@espci.psl.eu}
\date{\today}
\begin{abstract} 
Based on the analysis of probability flow, where the First Passage (FP) is realised as the sink of probability,
we summarise the protocol to find the distribution of the First Passage Time (FTP). We also describe the corresponding formula for the discrete time case. 
\end{abstract}
\maketitle

\section{Introduction} 
As the title shows, this note aims at providing with a concise r\'esum\'e of the protocol and its derivation of the first passage time (FTP) distributions on the discrete Markov network whose transition rates are constant in time.  While the well-written reviews and books on the derivation of the {\it mean} FPT are available for physicists 
\cite{hanngi,FTP-Redner2001}, the description of the probability {\it distribution} of the FTP 
\cite{Haake-Glauber-FPT1981} is not easily findable in the reviewed articles or books at least for the author. Nevertheless the FTP distribution is often useful when we analyse the transition network focusing on some specific states or transitions from which we extract, for example, the  entropy production \cite{Edgar-QM-stopping-time-PRL2019,Neri-stopping-time-PRL2020}.

The note originates from the course note of the author for the graduate course students. The contents may have been known among the experts of probability theory. Nevertheless, the note is presented here since the author had quite a few requests to bring it publicly accessible so that the users can cite it instead of explaining from scratch in their articles.

Below, after the definition of the problem and the introduction of the notations
(\S \ref{sec:intro}), we derive the FTP distribution in the main part (\S \ref{sec:distribution}). 
We also give a concrete example as an appendix \ref{sec:example}, which shows how the {\it reduced} master equation works.
A completely parallel formalism is also given for the FTP distribution in the discrete time problem (\S \ref{sec:discrete}). The possibility of generalisation is discussed in \S \ref{sec:conclusion}.

\section{Master Equation and the First Passage Time (FTP) Problem}\label{sec:intro}
We recall the master equation on the discrete network and introduce some notations.
Some problems are solvable much more easily by an ensemble approach, rather than focusing on individual realisations. The master equation is a basic tool for this approach.Those who are familiar to these notions may jump to the next section.

\subsection{Master equation} 
Let us denote by $P_\alpha(t)$  the probability that the system is in the state $\alpha$  at time $t.$
We will use later the vector notation $\vec{P}(t)$ to represent the all components $\{P_\alpha(t)\}.$ 
Up to the precision of $\mathcal{O}(dt)$ the change of this probability is
\beqa
P_\alpha(t+dt)-P_\alpha(t)
 &&=-\inSbracket{\sum_{\beta(\neq \alpha)} w_{\beta\leftarrow\alpha} dt} \,P_\alpha(t)
\cr &&+\sum_{\beta(\neq \alpha)} {w_{\alpha\leftarrow\beta}dt\,P_\beta(t)}.
\eeqa
In dividing by $dt$ and rearranging the terms, it can be cast in the vector-matrix form of 
the { master equation}:
\[\frac{d}{dt}P_\alpha(t)=\sum_\beta M_{\alpha,\beta}P_{\beta}(t), \]
or 
$$\frac{d}{dt}\vec{P}(t)={\sf  M} \vec{P}(t), $$
where  $M_{\alpha,\alpha}:=-\sum_{\gamma (\neq \alpha)}w_{\gamma\leftarrow\alpha}$
and $M_{\alpha,\beta}:=w_{\alpha\leftarrow\beta}$ for $\alpha\neq\beta.$ Then $\sf M$ is a square matrix.  
All the off-diagonal components are non-negative while the diagonal components are non-positive, such that $\sum_\alpha M_{\alpha,\beta}=0.$
We focus on the case where $\sf M$ is {\it independent of time}.
Then  the solution for the initial value problem reads 
$\vec{P}(t)=e^{{\sf M}t}\vec{P}(0).$%
\footnote{The exponential of a matrix $e^{\sf R} $ is defined by $e^{\sf R}:=\sum_{n=0}^\infty \inv{n!}{\sf R}^n.$}

\subsection{First-passage time (FPT) problem}
\begin{figure}[h]  \begin{center} 
\includegraphics[width=5cm,angle=-0]{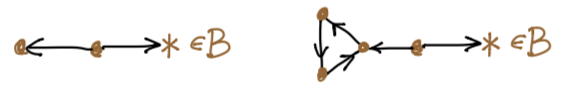} 
\caption{The cases where $\hattau_{\tiny \rm FP}(\alpha_0)=\infty$ 
can happen. We exclude such cases from our consideration.
 \label{fig:infFT}}
\end{center} \end{figure}

 At the initial time, $t=0,$ the system is put in the state $\alpha_0$.\footnote{The extension to the probabilistic initial condition is straightforward.}
An individual realization allows to make jumps from a state to the other. 
When the system arrives for the first time at one of the ``goal states'' named $B=\{\beta_i\}$
the stochastic process {\it stops}.\footnote{In the network language, the transition from node to node on the network end once the state jumps to one of the nodes belonging to $B$.}
We denote by $B^c$ those states which are complement of $B.$ 
  Our interests is in the statistics of the time and the last transition into $B$ at the first passage.
 We denote by $\hattau_{\tiny \rm FP}(\alpha_0)$ this  time of the first arrival.\footnote{\null{When a more precision is required, we define that the  $\tau$ is such that the system is found in $B^c$ through $0\le t\le\tau$ but in $B$ for $t>\tau.$}}
This is called the {\it first passage time} (FPT). This is a random variable.
If $\alpha_0\in B,$ we define $\hattau_{\tiny \rm FP}(\alpha_0)=0.$  
Hereafter, we suppose that $\alpha_0\in B^c.$ 
In general it can happen that $\hattau_{\tiny \rm FP}(\alpha_0)=\infty;$  when the network contains any ``dead-end'' state other than $B,$ the system's state may remain in $B^c$ forever, see Fig.~\ref{fig:infFT}.
Hereafter, we exclude such cases; we assume that from any state (node) $\alpha_0$ in $B^c$ the system can eventually reach one of the $B$ states.

\section{Derivation of the FTP distribution}\label{sec:distribution}
\subsection{Basic idea}
In the ensemble-based view, the master equation can be interpreted so that  a unit ``mass'' (in fact the probability being 1) is injected at the node $\alpha_0$ at $t=0,$ and then 
the injected mass (probability)  {\it flows} out through the links at the rate being the transition rate.

\begin{figure}[h]  \begin{center} 
\includegraphics[width=2.7cm,angle=-0]{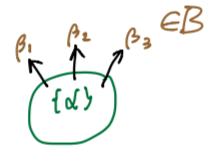} 
\caption{The arrows show the one-way exits to the goal sites,
$\{\beta_1,\beta_2,\beta_3\} \subset B,$ from the group of states, $\{\alpha\}=B^c,$ which are represented by a circle.
 \label{fig:exit1}}
\end{center} \end{figure}

For the purpose of analysing the FTP  we {modify} the network such that any state belonging to $B$ is replaced by a sink (``black hole (of mass)''), where no outward transition occurs, see Fig. \ref{fig:exit1}.
Then the probability $Prob(\hattau(\alpha_0)\in [t,t+dt])$ is given by the {\it cumulated flow} of the mass (probability) $B^c\to B$ during this time interval. When the states are indexed by  continuous parameters such as Euclidean coordinates, $B$ is usually represented as absorbing boundaries. All what remains to do is the formulation of this idea as a recipe.
\subsection{Basic recipe }\label{subsec:basic-recipe} 
Let us begin by an extremely simple network of which $B^c=\{1\}$ and $B=\{*\}$. The general case can be formulated later by using the analogy to this case.
The only reactions here are $1{\longleftrightarrow}*$, with the rates,
$w$ for $1\rightarrow *$ {\it and} $w'$ for $1\leftarrow *,$ respectively.
\ben
\item[Step 1.]  Among the normalized probabilities, $P_1$ and $P_*, $ 
we exclude $P_*,$  and omit the transition ($w'$) emitted from $*$.\\
Then we solve only { $\frac{d}{dt}{P}_1(t)=-\inv{\tau} {P}_1(t)$}  from $P_1(0)=1.$
We find {$P_1(t)=P_1(0)e^{-t/\tau}.$} 
\item[Step 2.] 
We understand that 
\beqa \label{eq:ftp1}
&& Prob\big(\mbox{\small ``the transition $1\to *$ takes place in the interval $[t,t+dt[$\,''}\big)
\cr && = P_1(t)-P_1(t+dt)\simeq - \frac{dP_1(t)}{dt}\, dt.
\eeqa 
Therefore,  $-\frac{dP_1(t)}{dt}$ is the {\it probability density} for the FPT to be $t$.
If we want to know $Prob\big(\mbox{``FPT is less than $t$''}),$ 
we can integrate; $\int_0^t(-\frac{dP_1(t')}{dt'})dt' =-P_1(t)+P_1(0)=1-P_1(t).$
It is intuitively correct because $P_1(t) =Prob\big(\mbox{``FPT is more than $t$''})$ in the present case.
\een

\subsection{Extended recipe}
 We study  the network allowing more than one inner states ($\in B^c$) and goal states ($\in B$).
\subsubsection*{Step 1. \rm Reduced transition network} 
We remove from the full transition network  
 all the transition edges {\it from} the group $B.$ --- We do this so that the process is over once the system reach the ``black hole'' $B$-states.
 These $B$-states then become just a sink, having only in-coming edges. As a consequence $P_\beta$'s with $\beta \in B$ appear no more.\\

We call the resulting network the {\it reduced transition network}.
\begin{figure}[h!!]  \begin{center} 
\includegraphics[scale=0.3,angle=-0]{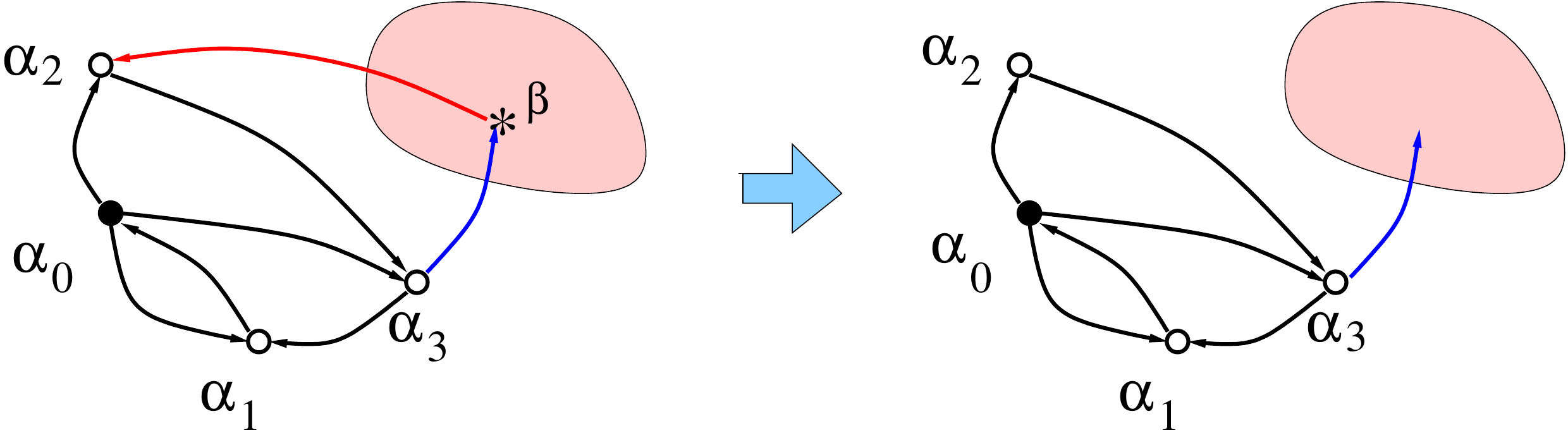} 
\caption{(Left) Original transition network including $B=\{\beta\}.$ (Right) Reduced network.
\label{fig:network1}}
\end{center} \end{figure}
Fig.~\ref{fig:network1} illustrates the construction of the reduced transition network. 

As the master equation, we are left with a {\it reduced master equation}, 
{$$\frac{d}{dt}\vec{P}^*(t)={\sf  M}^* \vec{P}^*(t),$$ }
where $\vec{P}^*(t)$ is the vector $\vec{P}(t)$ {\it less} the states in $B$-group, and the reduced matrix ${\sf M^*}$ is again square.\footnote{\label{foot:Mred} One can say that, if we write the original equations as ${\sf K}=\left( \begin{matrix} B^c\leftarrow B^c & B^c\leftarrow B  \\       B\leftarrow B^c  & B\leftarrow B  \\    \end{matrix}\right),$ then the reduced  matrix ${\sf M^*}$ contains only the block $ B^c\!\! \leftarrow \!\! B^c$. 
    }

The set of master equations for the original (i.e., non-reduced) network contains the equations $\frac{dP_{\beta}}{dt}=\ldots$ with $\beta\in B.$ In the reduced network, those equations are absent. 
The solution of the reduced master equation can be written as {$\vec{P}^*(t)=e^{{\sf M}^* t}\vec{P}^*(0).$}
If the initial state is $\alpha_0$ we write $\vec{P}^*(0)=|\alpha_0\ra.$
Comparing with the equations in \S~\ref{subsec:basic-recipe}, we see how the basic scheme has been extended.\\

\noi
{\it {}\,\,Numerically,} $e^{{\sf M}^*t}$ is calculated using the spectral decomposition,  
${\sf M}^*=\sum_\nu |\lambda_\nu\ra \lambda_\nu \la \lambda_\nu |$,
\footnote{ 
The  eigenvector \protect{$ | \lambda_\nu\ra$ }
and its conjugate, \protect{$\la \lambda_\nu |,$}
 have generally different components but can be made so that 
\protect{$ \la \lambda_\nu | \lambda_\mu\ra=\delta_{\nu\mu} $ }
 is assured. More can be  learned, check under the key word polar decomposition (of square matrix).
}
 that is, $e^{{\sf M}^* t}=
\sum_\nu |\lambda_\nu\ra e^{\lambda_\nu t} \la \lambda_\nu |.$
Once $e^{{\sf M}^* t}$ is obtained, $\vec{P}^*(t)$ is given by the matrix-vector product,
$\vec{P}^*(t)=e^{{\sf M}^* t}\vec{P}^*(0)=e^{{\sf M}^* t}|\alpha_0\ra.$ \\

\noi
{\it Remarks}: \\
(i) The matrix ${\sf M}$ or ${\sf M}^*$ can have complex eigenvalues. Nevertheless, when ${\sf M}$ or $e^{{\sf M}t}$ are applied to a physically meaningful $\vec{P}^*,$ the result is always physically meaningful. \\
(ii) Having excluded  the diverging FPT, all the eigenvalues of ${\sf M}^*$ must have strictly negative real part because $\vec{P}^*(t)$ with any initial $\alpha_0$ should decay to the reduced zero-vector, $\vec{0}^*,$ for $t\to\infty.$\footnote{cf. The {\it non}-reduced $\sf M$ must have at least a zero eigenvalue.}
\\
{(iii)} The evolution of $\vec{P}^*(t)$ is generally different from 
the $B^c$-part of the full evolution, $\vec{P}(t),$ because the former excludes the ``returning from $B$-group.''

\subsubsection*{Step 2. \rm Probability distribution of FPT}
In order to simplify the notation, we introduce a special row vector, \`a la Dirac, in the reduced state space, $\la \, |$ whose $\alpha (\in{B^c})$ components are all unity (=1).\footnote{More precisely we should write \protect{$|{\alpha_0}^*\ra$} or \protect{$\la\,^* |$} etc. for the reduced network. We will, however, omit \protect{$^*$} whenever it is clear.} 
With $\la \, |$  we have  the short hand like $\sum_{\alpha \in B^c} P^*_\alpha= \la \, | \vec{P}^*\ra,$  or  $\sum_{\alpha \in B^c} {M}^*_{\alpha,\mu}= \la\,| {\sf M}^*|\mu\ra.$\footnote{cf. In the non-reduced network \protect{$\la \,|{\sf M}=0.$}}\\

$\la \, | \vec{P}^*(t)\ra$ is the probability that the system has not reached any goal state at time $t.$ Then  probability that the system reaches for the first time one of the $B$ states in the interval
$[t,t+dt[$ reads, in analogy to (\ref{eq:ftp1}), 
\beqa \label{eq:FTP}
 Prob\inRbracket{\hat{\tau}_{\tiny \rm FP}(\alpha_0)\in [t,t+dt[}
 &=&
\la\,|\vec{P}^*(t)\ra-\la\,|\vec{P}^*(t+dt)\ra
\cr &=& -\la\,|\frac{d\vec{P}^*(t)}{dt}\ra\, dt
\cr &=& -\la\,|{\sf M}^* |\vec{P}^*(t)\ra \, dt.
\eeqa 
The {\it probability density of FTP} is, therefore, the coefficient of $dt$ on the right extreme part :
$$ 
p_{FPT}(t)= -\la\,|{\sf M}^* |\vec{P}^*(t)\ra
$$
The normalisation is $\intOI p_{FPT}(t)dt=1.$\\
{\it {}\,\,Numerically}: As we already have $|\vec{P}^*(t)\ra,$ the further matrix-vector 
product with $( -\la\,|{\sf M}^*) $ from the left gives $p_{FTP}(t).$\\

\noi Theoretically :  We can rewrite the r.h.s. of (\ref{eq:FTP}) to represent:
\beq \label{eq:FPT-compo}
 Prob\inRbracket{\hat{\tau}_{\tiny \rm FP}(\alpha_0)\in [t,t+dt]}=\sum_{\beta\in B}\sum_{\alpha\in B^c}w_{\beta\leftarrow\alpha} P_{\alpha}^*(t)dt,
\eeq
where only those $\alpha\in B^c$ that have non-zero rate $w_{\beta\leftarrow \alpha}$ to any goal state $\in B$ contribute. (Note that  $w_{\beta_i\leftarrow\alpha}$ is not an element of ${\sf M}^*.$)
--- We propose two versions of intuitive explanation of (\ref{eq:FPT-compo}).\\
(Version1) 
The surviving probability, $\la\,|P^*(t)\ra,$ decays {\it only through} the transitions $w_{\beta\leftarrow\alpha}$ ($\beta\in B, \alpha\in B^c$). We, therefore, measure the escaping flux of probability  through those channels. \\
(Version 2) 
The reduced master equation has the divergence form, $d|P^*\ra/dt=-\nabla\cdot J.$ 
Then its ``volume integral'' $\la\, |$ can be rewritten by the {\it ``Gauss' divergence theorem,''}  $\int_V dV (\nabla\cdot J)=\oint_{\partial V} d\vec{S}\cdot J.$ 
Thus $-\frac{d\la\, |P^*\ra}{dt}= \oint_{\partial V} d\vec{S}\cdot J$ is found.
Lastly the $J$ on the boundary $\partial V$ is the ``fluxes to $B$'' through $w_{\beta\leftarrow\alpha}$ ($\beta\in B, \alpha\in B^c$).\\


\subsection{Outcomes of FTP distribution}
\subsubsection{Mean FPT (MFPT)}
Often we focus on the mean FTP. By definition,
$$E[\hat{\tau}_{\tiny \rm FP}|\alpha_0]= \int_0^\infty t\,Prob\inRbracket{\hat{\tau}_{\tiny \rm FP}(\alpha_0)\in [t,t+dt]}.$$
Using (\ref{eq:FTP}) and the { integration by parts} w.r.t. time $t$, we have
\beqa \label{eq:MFPT}
E[\hat{\tau}_{\tiny \rm FP}|\alpha_0] 
&=& \int_0^\infty t\, \inRbracket{ -\la\,|\frac{d\vec{P}^*(t)}{dt}\ra}\, dt
\cr &=&
\int_0^\infty \la\,|P^*(t)\ra dt=\la\, |\inRbracket{-\inv{{\sf M}^*}}| \alpha_0\ra.
\eeqa
To have the last equality, we used $|P^*(t)\ra=e^{{\sf M}^* t}|\alpha_0\ra.$
Many books for physicists mentions only the MFPT, because the calculation of 
MFPT does not require the FTP distribution. As we have seen, however, 
the FPT distribution is simpler and more basic. \\

\noi
{\it {}\,\,Numerically}:
The r.h.s. of the MFPT expression requires the calculation of the inverse of ${\sf M}^*.$\footnote{\protect{${\sf M}^*$} is invertible. cf. \protect{${\sf M}$} has zero eigenvalue and therefore non-invertible.}
A usual protocol, instead, is to multiply by $M^*_{\alpha_0,\alpha}$ and take the sum over $\alpha_0$ over the reduced network states. Then we have\footnote{In the vector-matrix sense, we multiply each side of the vector equation (\ref{eq:MFPT}) the transposed matrix, $({\sf M}^*)^t,$ from the left. (Attention: ${\sf M}^*$ is {\it not} the transpose of $\sf M.$)}
\beqa \label{eq:MFPTbis}
\sum_{\alpha_0} E[\hattau|\alpha_0]\, M^*_{\alpha_0,\alpha}
&=&-\sum_{\alpha'}\sum_{\alpha_0}
(\inv{{\sf M}^*})_{\alpha',\alpha_0}M^*_{\alpha_0,\alpha}
\cr &=&-\sum_{\alpha'}\delta_{\alpha',\alpha}
\cr &=& -1
\qquad \mbox{for  $\forall \alpha.$}
\eeqa
By using the solver of coupled linear equations we find $E[\hattau|\alpha_0]$ for $\forall \alpha_0$.

\subsubsection{Exit problem}
Sometimes we are not interested in  the value of FPT, but rather interested in how it finished. That is,  among the goal states $B$,
we ask which $\beta$ has absorbed the state point. See Fig. \ref{fig:exit1}. 
For this purpose, we segregate the result (\ref{eq:FPT-compo}). If we like to know the probability that the state finish in $\beta_i (\in B),$ we calculate the partial probability:
\beqa
\frac{Prob\inRbracket{\mbox{exit to $\beta_i$}|\alpha_0}}{
Prob\inRbracket{\mbox{exit to $B$}|\alpha_0}}
&=&
{\int_0^\infty  \sum_{\alpha\in{B^c}}w_{\beta_i\leftarrow\alpha}  P_{\alpha}^*(t)dt}
\cr &=&
{\sum_{\alpha\in{B^c}}w_{\beta_i\leftarrow\alpha} \la\alpha |\inRbracket{-\inv{{\sf M}^*}}| \alpha_0\ra},
\cr &&
\eeqa
where we have noticed our setup, $Prob\inRbracket{\mbox{exit to $B$}|\alpha_0}= 1$ \footnote{When \protect{$Prob\inRbracket{\mbox{exit to $B$}|\alpha_0}\neq 1,$} i.e., in the presence of internal trapping, we should divide the second and the last expressions, respectively, by this quantity, i.e. \protect{${
{\int_0^\infty \sum_{\beta\in B} \sum_{\alpha\in{B^c}}w_{\beta\leftarrow\alpha}  P_{\alpha}^*(t)dt} }=$}  \protect{${\sum_{\beta\in B}\sum_{\alpha\in{B^c}}w_{\beta\leftarrow\alpha} \la\alpha |\inRbracket{-\inv{{\sf M}^*}}| \alpha_0\ra}.$} }. As noticed above  $w_{\beta_i\leftarrow\alpha}$ is not an element of ${\sf M}^*.$

 The exit problem plays an important role in the evolution, namely the fixation of a new genotype among the (finite) population. Either the extinction of  the new ( and overwhelmingly neutral) genotypes or the extinction of the original one are the two exits.

\section{Discrete time version}\label{sec:discrete}
\paragraph{"Master equation"}:
We denote by $n\in \N_0$ the discrete time and $\vec{P}(n)$ denotes the probability vector
of the original (non-reduced) network.
The evolution of $\vec{P}(n)$ writes
$$\vec{P}(n+1)={\sf K} \vec{P}(n);$$
where the transfer matrix ${\sf K}$ is supposed to be constant in time.
The normalization of the probability imposes $\la \,|{\sf K}=\la\,|,$ or 
$\sum_j K_{j\leftarrow i}=1$ for $\forall i.$
Again we suppose the case when any states in $B^c$ can eventually reach one of the $B$ states.

\paragraph{First Passage Time}: 
When the initial state $\alpha_0$ is already in $B$ block, we define that FTP is zero.
For $\alpha_0 \in B^c$ the FPT,  $\hat{\tau}_{FP}(\alpha_0),$ should be positive. It is, therefore, consistent to say $\hattau_{FP}(\alpha_0)=1$ if the system enters  $B$ at $t=1.$
In general we say $\hattau_{FP}(\alpha_0)=n$ if the system enters  $B$ at $t=n$ but have remained in $B^c$ for $0\le t< n.$

\paragraph{``Reduced transfer matrix"}: 
We introduce the state space that contains only the states belonging to $B^c,$ and 
also introduce the reduced transfer matrix ${\sf K}^*$ that lacks the rows and lines corresponding to $B$ states.\footnote{As was discussed in the footnote [10],  
if we divide ${\sf K}$ into four blocks symbolically, ${\sf K}=\left( \begin{matrix} B^c\leftarrow B^c & B^c\leftarrow B  \\       B\leftarrow B^c  & B\leftarrow B  \\    \end{matrix}\right),$  the reduced one ${\sf K}^*$ is the  $B^c\leftarrow B^c$ square block.}
Recall that the element $K_{\beta\leftarrow\alpha}$   with 
$\alpha \in B^c$ and $\beta\in B$ is contained in ${\sf K}^*$ in the form, $K_{\alpha \leftarrow \alpha}= 1-\sum_{\beta\in B}K_{\beta \leftarrow \alpha}.$\footnote{By contrast the elements like $K_{\alpha \leftarrow \beta}$ with $\alpha\in B^c$ and $\beta\in B$ appear only in the blocks $B^c\leftarrow B$ and $B\leftarrow B.$}

\paragraph{Results}: 
Because the basic idea is the same as the case of continuous time, we only write some resultant formulas. We shall denote by $S_n$ the state of the system at time $n(\ge 0).$
We suppose that $\alpha_0\in B^c.$ 
:\\
\noi i) Cumulated probability of FPT for $n\ge 0$: 
\beq
Prob(\hattau_{FP}> n|  S_{n=0}=\alpha_0 )
 = \la \,|({\sf K}^*)^n |\alpha_0\ra. 
\eeq
Especially $Prob(\hattau_{FP}> 0|  S_{n=0}=\alpha_0 )=1.$

\noi ii) Probability of FPT being $n$: \\
In our setup, $Prob(\hattau_{FP}= 0| S_{n=0}=\alpha_0 ) = 0,$ and 
for $n\geq 1$
\beqa
&&Prob(\hattau_{FP}= n| S_{n=0}=\alpha_0 ) 
\cr &&=
Prob(\hattau_{FP}>\! n-1| S_{n=0}=\alpha_0 )\!
 -\! Prob(\hattau_{FP}>\! n| S_{n=0}=\alpha_0 ) 
\cr &&=  \la \,|({\bf 1}-{\sf K}^*)({\sf K}^*)^{n-1} |\alpha_0\ra. 
\eeqa
Using $\la \, |\alpha_0\ra =1$ and $({\sf K}^*)^\infty={\sf 0}$ 
the normalisation reads
\beq \sum_{n=1}^\infty Prob(\hattau_{FP}= n| S_{n=0}=\alpha_0 ) 
= \la \, |\alpha_0\ra - \la\, |({\sf K}^*)^\infty |\alpha_0\ra =1,
\eeq

\noi iii) Probability of FPT with specified route from 
$\alpha_s\in B^c$ to $\beta_s\in B.$: 
\beqa
&&
Prob(\hattau_{FP}= n\wedge \mbox{(through $\beta_s\leftarrow \alpha_s$)} | S_{n=0}=\alpha_0 ) 
\cr &&= {\sf K}_{\beta_s\leftarrow\alpha_s} 
\la \alpha_s|({\sf K}^*)^{n-1}   |\alpha_0\ra,
\eeqa
because $\la \alpha_s|({\sf K}^*)^{n-1}   |\alpha_0\ra$ gives the probability of finding the system in $\alpha_s$ after $(n-1)$ steps then we multiply the conditional probability of the specific exit, ${\sf K}_{\beta_s\leftarrow\alpha_s}.$ (Note that ``${\sf K}^*_{\beta_s\leftarrow\alpha_s}$'' does not exist.)
The normalisation should be such that 
\beqa
&&\sum_{\beta_s\in B}\sum_{\alpha_c\in B^c} 
Prob(\hattau_{FP}= n\wedge \mbox{(through $\beta_s\leftarrow \alpha_s$)} | S_{n=0}=\alpha_0 ) \cr &&=Prob(\hattau_{FP}= n| S_{n=0}=\alpha_0 ). 
\eeqa

\section{Concluding discussion}\label{sec:conclusion}
\begin{figure}[h!!]
\centering
\vspace{-0.5cm}
\subfigure[\null] 
{\label{fig:sublabel1}    \includegraphics[width=2cm]{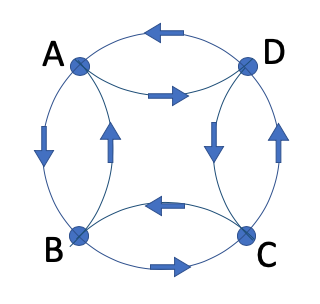}} 
\hspace{0.5cm}
\subfigure[\null] 
{\label{fig:sublabel2}    \includegraphics[width=5cm]{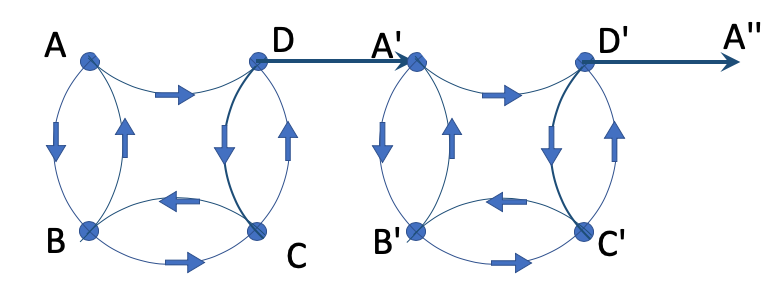}} 
\hspace{0.5cm}
\subfigure[\null] 
{\label{fig:sublabel3}    \includegraphics[width=3.cm]{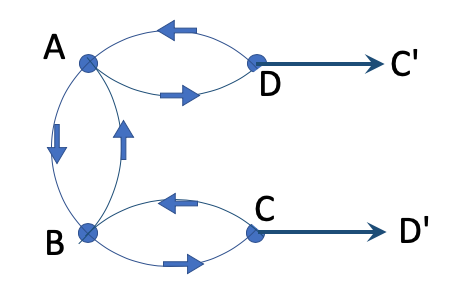}} 
\hspace{0.5cm}
\subfigure[\null] 
{\label{fig:sublabel4}    \includegraphics[width=3.7cm]{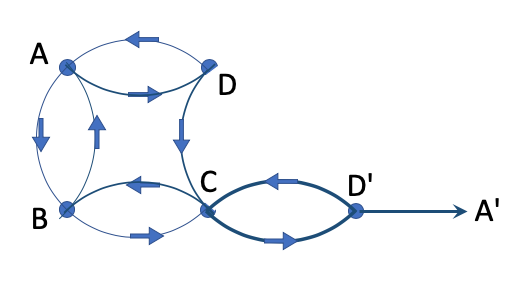}} 
\caption{See the main text.} 
\label{fig:conclusion} 
\end{figure}

\begin{figure}[h!!]
\centering
\vspace{-0.5cm}
\subfigure[\null] 
{\label{fig:sublabel5}    \includegraphics[width=3cm]{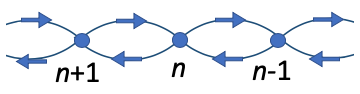}} 
\hspace{0.5cm}
\subfigure[\null] 
{\label{fig:sublabel6}    \includegraphics[width=3cm]{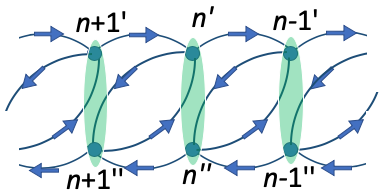}} 
\hspace{0.5cm}
\subfigure[\null] 
{\label{fig:sublabel7}    \includegraphics[width=3.cm]{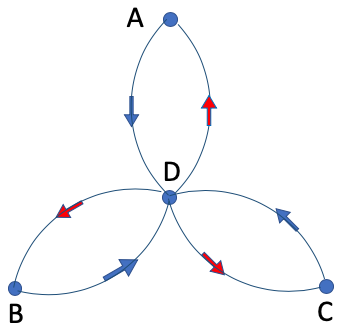}} 
\hspace{0.5cm}
\subfigure[\null] 
{\label{fig:sublabel8}    \includegraphics[width=3.cm]{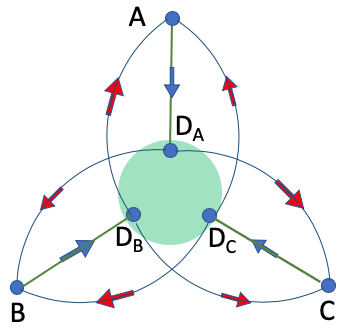}} 
\caption{See the main text.} 
\label{fig:conclusion2} 
\end{figure}

Based on the analysis of probability flow, where the FPT is realised as the sink of probability, we have summarised the protocol to construct the FPT distribution. 
Key is to reform the transition network in the way that the goal states are made to be the sink, even valid for non-Markovian case.  The reduced transition rate matrix ${\sf M^*}$ is 
a partial diagonal block of the original one, ${\sf M},$ in the simple cases that are treated in the main text.
  However, we don't stress this view ``reduced'' too much because the idea of absorbing nodes (sinks) is more generally applicable; we can sometimes {\it add}  nodes or replicate the original network so as to adapt to more advanced FP problem. For example, we can replace the particular links $\alpha_1\stackrel{M_{12}}{\to}  \alpha_2$ and $\alpha_2\stackrel{M_{21}}{\to} \alpha_1$ in a network by       $\alpha_1\stackrel{M_{12}}{\to} *$ and $\alpha_2\stackrel{M_{21}}{\to} *,$ respectively, where $*\in B$ is the added absorbing node.  In the ring network \cite{Pedro-FPT-arXiv} depicted by 
  Fig.\ref{fig:sublabel1}, Figs.\ref{fig:sublabel2}, \ref{fig:sublabel3} and \ref{fig:sublabel4} implement, respectively, 
  (b) the {\it counter} of the number of $D\to A$ transitions ($D\to A'$ giving the first one),
  (c) the {\it competition} of the earlier passage of $D\to C$ vs $C\to D,$ and 
  (d) the first {\it consecutive} transitions $C\to D\to A.$ 
Moreover, if we extend the idea of replicating the nodes, we can {\it Markovianize} exactly some non-Markovian model having a two neighbors (Fig.\ref{fig:sublabel5}$\to$(b)) or three neighbors (Fig.\ref{fig:sublabel7}$\to$(d)). In both cases the node memories from where it came.

\appendix
\section{A paradoxical example of  MFPT}\label{sec:example}
We apply the present approach for the continuous time to find a MFPT of a particular example which is somehow counter-intuitive, the situation being related to so called {\it waiting-time paradox}.

\begin{figure}[h]  \begin{center} 
\includegraphics[scale=0.4,angle=-90]{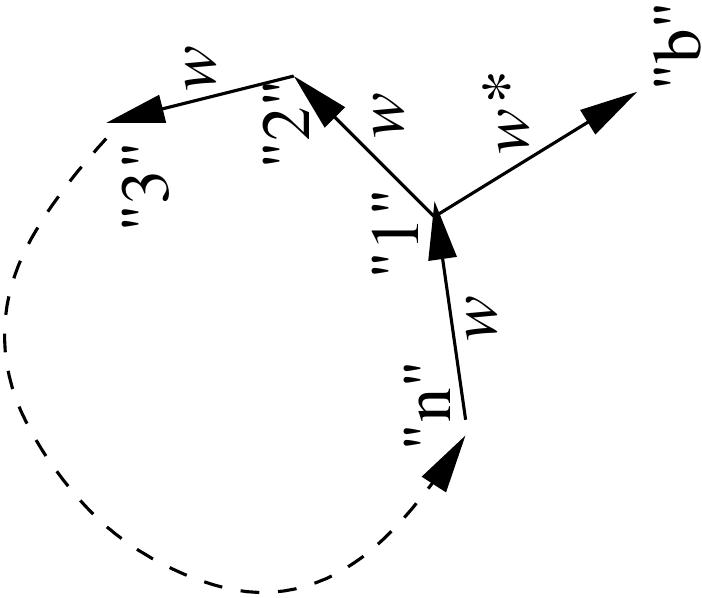} 
\caption{The circular transition network with an exit. \label{fig:network3}}
\end{center} \end{figure}

Let us consider a circular transition network with unique exit like Fig.~\ref{fig:network3}. The network is a loop with $n$ nodes complemented by a unique exit from ``$1$'' to ``$b$''. The uni-directional transition rate on the loop is uniform (rate $w$) and the transition rate to the exit is $w^*$.
Starting at $t=0$ from the state ``$\alpha_0$" among ``1'' to ``$n$'', we like to know the expectation value of the  FPT ($E[\hat{\tau}_{\tiny \rm FP}|\alpha_0]$), which we denote by $\tau_{MFP}(\alpha_0).$ 
\mbox{}\\

\noi
\paragraph{Solving for MFPT}
\ben
\item We write down the master equation for $p_1,\ldots, p_n,$ in the form like $dp_j(t)/dt=\ldots$ with $1\le j\le n.$
Then from this equation, find the reduced matrix ${\sf M}^*$.
This matrix reads  (The diagonal elements are negative.)
\[{\sf M}^*=\left(  
\begin{array}{ccccc}
{-(w+w^*)}  & 0  & \ldots   & 0 & w\\
w  & {-w}  & 0  & \ldots & 0\\
0  & w  & {-w}  & \ldots & 0\\
0    & \ldots  & w  &\ldots & 0\\
\ldots    & \ldots  & \ldots  &\ldots & \ldots\\
0    & \ldots  & w  &{-w}  & 0\\
0  & \ldots  & \ldots  & w & {-w}
\end{array}
\right)
\]

\item 
We then write down the coupled linear equations for $E[\hat{\tau}_{\tiny \rm FP}|\alpha_0],$
see (\ref{eq:MFPT}). In the present case,\\
 $(\tau_{MFP}(1), \tau_{MFP}(2),\ldots, \tau_{MFP}(n)).{\sf M}^*=(-1,-1,\ldots, -1)$ gives
\beqa
-\tau_1 \frac{w+w^*}{w}+\tau_2&=& -\inv{w},
\cr -\tau_i+\tau_{i+1}&=&-\inv{w}\, \qquad (2\le i\le n-1),
\cr \tau_1-\tau_n&=& -\inv{w},
\eeqa
where $\tau_i\equiv \tau_{MFP}(i).$ 
Therefore, the solution is\footnote{The details:  By adding the middle equations  we find $\tau_n-\tau_2=-\frac{n-2}{w}.$ Next, substituting this $\tau_2$ into the first equation, we have $\tau_n-\tau_1\frac{w+w^*}{w}=-\frac{n-1}{w}.$  Finally, substituting this $\tau_n$ into the last equation, we have  $-\tau_1\frac{w^*}{w}=-\frac{n}{w}.$  } 
$\tau_1=\tau_{MFP}(1)=\frac{n}{w^*}$ and the others  are 
$\tau_i=\tau_1+\frac{n-i+1}{w},$  ($2\le i \le n$).\\
\een

{Surprisingly the result is independent of $w$.}
We like to discuss qualitatively the result in the two extreme cases, i.e., $w^*/w\ll 1$ and $w^*/w\gg 1.$ 
\bde
\item[For $w^*/ w\ll 1$:] We can find $\tau_{MFP}(1)$ approximately as follows.
Because of the relatively high transition rate along the loop, the mean time for an entire circulation, $\frac{n}{w}$, is much shorter than the waiting time from ``1'' to ``b'', $\inv{w^*}.$ Therefore, the probability is almost evenly shared among the $n$ states along the loop, that is, $P_1\simeq \cdots \simeq P_n\simeq \frac{P_{loop}(t)}{n},$ where $P_{loop}(t)$ is the total probability on the loop (i.e. in the group $B^c$).
From such quasi-steady state, the probability flows out slowly to ``b''. The decay of 
$P_{loop}(t)$ reads $\frac{d}{dt}P_{loop}(t)=-w^* P_1(t)= -w^* \frac{P_{loop}(t)}{n}.$ We then have 
$P_{loop}(t)=e^{-\frac{w^*}{n}t}.$ Since the probability density of FPT is $-\frac{d}{dt}P_{loop}(t)= \frac{w^*}{n}P_{loop}(t),$ we find $E[\hat{\tau}_{\tiny \rm FP}|\mbox{``1''}]={\frac{n}{w^*}}.$

\item[For $w^*/ w\gg 1$] We can find $\tau_{MFP}(1)$ approximately as follows.\footnote{This case is more subtle. A simple-minded argument could gave $(n+1)/w^*$ approximately.}
 In most cases, the system will exit directly, after the time  $\sim 1/w^*$. The probability for such case is $w^*/(w+w^*)= 1-\mathcal{O}(\frac{w}{w^*}).$  However, in rare case with the probability, $w/(w+w^*)\simeq 
\frac{w}{w^*},$ the system jumps to ``2'' instead of ``b''. Let us denote by $\Delta t_{1\to 2}$ the time took from ``1'' to ``2'' . Once the system finds in ``2'', we should count time $\sim \frac{n-1}{w}$ until it returns to ``1''. Now if we count these two cases, the mean
FP time is estimated to be 
\beqa
&&E[\hat{\tau}_{\tiny \rm FP}|\mbox{``1''}]
\cr &&=[1-\mathcal{O}(\frac{w}{w^*})] \frac{1}{w^*}+\frac{w}{w^*}
\inRbracket{\Delta t_{1\to 2}+\frac{n-1}{w}}
\cr &&= {\frac{n}{w^*}} +\frac{w}{w^*}\Delta t_{1\to 2}.
\eeqa
The last point is the estimation of $\Delta t_{1\to 2}.$ If it were estimated to be \underline{$\simeq \inv{w}$}, we would have a wrong result, $\frac{n+1}{w^*}.$ The correct estimation of 
$\Delta t_{1\to 2}$ is $\frac{1}{w+w^*}\underline{\simeq \inv{w^*}}.$ 
This issue is related to the known paradox as explained below:
\paragraph{ Waiting-time paradox}:
Suppose a network given by $2\stackrel{w}{\longleftarrow}1\stackrel{w^*}{\longrightarrow}b$ and start by ``1'' at $t$ = $0.$ 
If we solve the {\it full} master equations for $(P_1,P_2, P_b)^t,$ we will have 
\beq
\inRbracket{\begin{matrix} 
P_1(t)\\P_2(t) \\ P_b(t)\\
\end{matrix}}
=\inRbracket{
 \begin{matrix} 
  e^{-(w+w^*)t}\\ \frac{w}{w+w^*}[1-e^{-(w+w^*)t}]\\
\frac{w^*}{w+w^*}[1-e^{-(w+w^*)t}]\\
\end{matrix}}
\eeq
Noting that $P_2(t)$ is the joint probability, $Prob(\hat{\tau}_{1\to 2}< t\,\wedge\, 1\to 2),$ on the one hand, and that the exit probability to $2$ is
$ Prob(1\to 2)=P_2(\infty)=\frac{w}{w+w^*}$ on the other hand,
we have 
\beqa 
&& Prob(\hat{\tau}_{1\to 2}< t\,|\, 1\to 2)
\cr &&=\frac{Prob(\hat{\tau}_{1\to 2}< t\,\wedge\, 1\to 2)}{Prob(1\to 2)}
\cr &&= 1-e^{-(w+w^*)t}
\eeqa
Therefore, when the transition $1\to 2$ takes place, it occurs in $[t,t+dt[$ 
at the probability, $\frac{d(1-e^{-(w+w^*)t})}{dt}dt =(w+w^*)e^{-(w+w^*)t} dt.$
In conclusion, when the transition $1\stackrel{w}{\to} 2$ competes with $1\stackrel{w^*}{\to} b,$ 
the mean waiting time of $1\to 2$ transition is shortened to $\inv{w+w^*}\simeq \inv{w^*}$ from $\inv{w}$ of the non-competing case.
\ede

\bibliographystyle{apsrev4-2.bst}
\bibliography{ken_LNP_sar}

\end{document}